\documentclass[12pt]{article}
\usepackage{amsmath}
\usepackage{latexsym}
\usepackage{bbm}
\usepackage{amssymb}

\def \vol {\text{vol}\,}
\def \slc {\text{SL}(2, \mathbb{C})}
\def\a{\alpha}

\def\b{\beta}

\def\d{\delta}

\def\e{\varepsilon}

\def\z{\zeta}

\def\th{\theta}

\def\l{\lambda}
\def\m{\mu}
\def\n{\nu}
\def\o{\omega}

\def\r{\rho}

\def\s{\sigma}

\def\D{\Delta}

\def\pa{\partial}

\def\half{\frac{1}{2}}

\def\tr{{\rm tr}}

\def\and{{\rm and}}

\def\IC{\mathbbm C}

\def\IZ{\mathbbm Z}

\def\A5{{AdS_5 \times S^5}}

\newcommand{\be}{\begin{equation}}
\newcommand{\bea}{\begin{eqnarray}}

\newcommand{\ee}{\end{equation}}
\newcommand{\eea}{\end{eqnarray}}

\begin{document}
\vspace*{-1.0in}
\thispagestyle{empty}
\begin{flushright}
CALT-TH-2020-001
\end{flushright}

\normalsize
\baselineskip = 18pt
\parskip = 6pt

\vspace{1.0in}

{\Large \begin{center}
{\bf M5-Brane Amplitudes}
\end{center}}

\vspace{.25in}

\begin{center}
John H. Schwarz\footnote{jhs@theory.caltech.edu}
\\
\emph{Walter Burke Institute for Theoretical Physics\\
California Institute of Technology 452-48\\ Pasadena, CA  91125, USA}
\end{center}
\vspace{.25in}

\begin{center}
\textbf{Abstract}
\end{center}
\begin{quotation}

The dynamics of a probe M5-brane, embedded as a hypersurface in eleven-dimensional
Minkowski spacetime, is described by a six-dimensional world-volume theory.
This theory has a variety of interesting symmetries some of which are obscure in the
Lagrangian formulation of the theory. However, as summarized in this review,
an alternative approach is to construct all of its on-shell tree-level scattering
amplitudes. This enables understanding all of the symmetries in a satisfying way.
This work is dedicated to the memory of Peter Freund.

\end{quotation}

\newpage

\pagenumbering{arabic}



\section{Introduction}

I am pleased to have an opportunity to contribute to this volume in memory of a
great physicist, my good friend Peter Freund. We never collaborated, but
I often enjoyed discussing physics with Peter,
and I admired his research. Reading his 2007 autobiographical reminiscences,
aptly entitled {\em A Passion for Discovery}, I learned about Peter's
courage as a student activist in Romania, where in 1956 he faced a row of tanks
that were aimed at him and his friends. It was fortunate for him, and the future of physics, that
the tanks did not fire and that
he was able to make his way to the United States. As it happens, my father was born in a nearby town in Romania that belonged to Hungary when he was there. The border was moved after
the first world war.

Peter's most famous and highly-cited paper, the 1980 ``Freund--Rubin" paper \cite{Freund:1980xh},
is only three pages long, yet it was remarkably insightful. It showed that 11-dimensional
supergravity has a maximally supersymmetric solution, involving four-form flux, in which the
11d geometry is $AdS_4 \times S^7$, a product of a 4d anti de Sitter space
and a 7d sphere. (There is also an
analogous solution of the form $AdS_7 \times S^4$.) Even though I found this result
interesting when it appeared, there were two reasons why I failed to fully appreciate it at that time.

The first reason for my failure to fully appreciate the Freund--Rubin solution
was that it appeared when Michael Green and I were developing superstring theory, which
requires ten dimensions. I was skeptical that 11d supergravity, which has severe UV
divergences, could have a quantum completion. On the other hand, I was sure that
superstring theory is UV finite. Therefore I mockingly described eleven dimensions as ``a 10\% error".
Of course, I now know better. The quantum completion of 11d supergravity even has a name -- M theory.
There might be a better formulation of M theory than is currently known,
but the existence of M theory as a consistent quantum theory is beyond question.

The second reason for my failure to fully appreciate the Freund--Rubin solution
was that I was interested in finding new
ways to compactify extra dimensions that can break supersymmetries and
leave four dimensions exactly or approximately flat while
the extra dimensions are compact and highly curved, so that they are unobservable at
sufficiently low energies.
The Freund--Rubin solution requires that the anti de Sitter and sphere curvatures
are comparable, which is certainly not realistic. Despite this misgiving,
three years later, when I was studying type IIB supergravity -- the low-energy
effective description of type IIB superstring theory -- Nick Warner and I realized that this theory
has an $AdS_5 \times S^5$ solution that is analogous to the Freund--Rubin solution.
This fact is mentioned for the first time in the concluding paragraph of \cite{Schwarz:1983qr}.

The profound importance of such anti de Sitter solutions of superstring theory
and M theory became clear to the
theory community with Maldacena's famous 1997 paper \cite{Maldacena:1997re}
proposing that such solutions have holographic dual descriptions
as conformal field theories (CFTs), which can be regarded as residing at the boundary
(at spatial infinity) of the anti de Sitter space.

\section{D-branes and M-branes}

The 3d CFT that is dual to the Freund--Rubin solution of M theory is associated
to a stack of M2-branes, which are half-BPS extended objects (with two spatial dimensions)
that exist in M theory. Similarly,
the 6d CFT that is dual to the $AdS_7 \times S^4$ solution of M theory
is given by a stack of M5-branes and the 4d
CFT that is dual to the $AdS_5 \times S^5$ solution of type IIB superstring theory
is given by a stack of D3-branes. The latter CFT,
which is the one that has been studied in the greatest detail, is
${\cal N} = 4 $ super Yang--Mills theory \cite{Brink:1976bc}. The rank $N$ of the $U(N)$
gauge group corresponds to the number of units of 5-form flux threading the $S^5$
(and the $AdS_5$). In each case, $N$, the number
of branes in the stack, is also the number of units of flux in the dual configuration.
Indeed, the branes are the sources for such flux.
The study of these three and other examples of AdS/CFT duality has
been a major theme of theoretical physics for more than two decades.

D-branes and M-branes are also of interest from other perspectives.
In particular, the world-volume theory of a single flat brane
can be studied as a probe of M theory or superstring theory. In this setting, the
word ``probe'' refers to an approximation in which the back reaction of the
brane on the M theory or string theory background is
neglected. This type of approximation is usually taken for granted in particle physics.
The relevant world-volume theories of such probe branes are
not conformal.\footnote{However, if the brane is embedded in an anti de Sitter space and localized
at a point in the other dimensions, then its world-volume theory has spontaneously
broken conformal symmetry.} Rather, they are extensions of
Born--Infeld theory, which is a nonlinear extension of Maxwell theory whose
Lagrangian has the form $L \sim \sqrt{- \det (g_{\m\n} + k F_{\m\n})}$.
Even though such theories probably do not have
a quantum completion without including additional massless degrees of freedom,
they are intriguing classical field theories with lots of symmetry and other special features
that are useful for various purposes.

There are two basic approaches to describing these theories. One is to formulate the Lagrangian
or, equivalently, the corresponding equations of motion.  In the mid-1990s my students and I expended
considerable effort formulating
D-brane and M5-brane Lagrangians, as did others. The Lagrangians for probe D-branes
embedded in 10d Minkowski spacetime are precisely
of the Born--Infeld type. In addition to the $U(1)$ gauge field of Born--Infeld theory,
they incorporate additional bosonic and fermionic degrees of freedoms that, together with
the $U(1)$ gauge field, comprise
maximally supersymmetric vector multiplets. These are
precisely the same kinds of supermultiplets that appear
in maximally supersymmetric Yang--Mills theories.

The 6d M5-brane theory is also DBI-like even though it involves a two-form tensor field, with a
self-dual field strength, instead of a Maxwell field. This type of field is quite awkward to incorporate
in a Lagrangian description, though various ways to do so have been developed
\cite{Perry:1996mk}\cite{Pasti:1997gx}\cite{Aganagic:1997zq}.
Despite the considerable effort expended in finding the formula for the Lagrangian,
I think it is fair to say that it is not very illuminating.
On dimensional reduction to 5d the self-dual tensor can be replaced by a vector,
by means of a duality transformation, and then the M5-brane theory becomes the D4-brane theory.
This procedure is sometimes called ``double dimensional reduction", because the 11d space is
simultaneously reduced to 10d. The D-brane Lagrangians are less awkward than the M5-brane one,
because they do not involve self-dual tensors.

A method of exploring properties of field theories that has become popular
in recent years is to present formulas for all of their on-shell scattering amplitudes.
Of course, this is the approach to the study of string theory that was utilized from its beginning
in 1968, building on the S-matrix program that had been developed in the preceding decade,
which is what I had been raised on in graduate school. Many
clever methods have been developed in the past decade for constructing amplitudes more
efficiently than by Feynman diagrams.
Some of them have been inspired by string theory \cite{Witten:2003nn}.
In some cases these amplitudes are given by remarkably elegant formulas.
My research during the last few years has focused on such field theory amplitudes.
Since the case of four dimensions appeared to be already in pretty good shape, I decided to focus on
6d supersymmetric field theories. I would prefer to be studying ten or eleven dimensions,
but the spinor-helicity methods that my collaborators and I have used, is not easily applied
to those cases. The hope is that six dimensions is a useful step in that direction. It may
be possible to extend spinor-helicity methods to ten or eleven dimensions. In fact, 
there are already some intriguing proposals \cite{Geyer:2019ayz}\cite{Bandos:2019zqp}
that I would like to understand better.

The plan for the remainder of this paper is to summarize what my collaborators and I have learned
about M5-brane scattering amplitudes \cite{Heydeman:2017yww}\cite{Cachazo:2018hqa}
\cite{Heydeman:2018dje}\cite{Schwarz:2019aat}.
Specifically, we have derived explicit formulas for the on-shell $n$-particle tree superamplitudes
of the M5-brane theory, which will be presented here. We have also obtained the amplitudes for
several other supersymmetric 6d theories, but I will not discuss them in this paper.
The M5-brane formulas describe the scattering of $n$ massless particles in 6d, each of which
belongs to a single tensor supermultiplet with $(2,0)$ supersymmetry. The formulas for these amplitudes
are quite concise and they have a lot of symmetry, most of which is made manifest. For one thing, the
amplitudes vanish unless $n$ is even. This $\IZ_2$ symmetry is generic for all Dirac--Born--Infeld
(DBI)-like theories.

The $n$-particle tree superamplitude ${\cal A}_n(\l, \eta)$ will be presented in the form of an
integral of an expression that contains a product of bosonic and fermionic
delta functions. Carrying out the integrations is an algebraic problem, whose solution
gives a rational function on the supermanifold times six momentum-conservation delta
functions and eight fermionic delta functions that describe conservation of half of the
supercharges. The supermanifold is parametrized by bosonic spinor-helicity
coordinates $\l_{ia}^A$ and Grassmann coordinates $\eta^I_{ia}$, where $i=1,2,\ldots, n$
labels the particles. The other indices will be described later.
One goal of the work that I am reviewing was to explore
the extent to which we can determine these rational functions from symmetry
considerations. Perhaps the lessons we have learned can be utilized
in the study of other interesting theories. Since symmetries are central to our work,
the plan is to discuss them in some detail before describing the amplitudes
themselves in the remainder of this paper.

\section{Symmetries of M5-brane superamplitudes}

As we have said, we are interested in studying the effective 6d field theory
associated to a flat probe M5-brane embedded in 11d Minkowski spacetime.
Before describing the tree-level scattering amplitudes of this field theory,
it is useful to understand their symmetries.

\subsection{Total permutation symmetry of the ${\mathbf n}$ scattered superparticles}

An entire supermultiplet, which consists of eight bosonic and eight fermionic
degrees of freedom, is represented by a scalar function of four Grassmann coordinates, $\eta^I_{a}$. The superamplitude, describing all possible $n$-particle scattering amplitudes of the M5-brane theory,
requires one such multiplet for each of the particles, $\eta^I_{ia}$, $i=1,2,\dots,n$.
Total permutation symmetry, $S_n$, of ${\cal A}_n(\l_i, \eta_i)$
ensures that Bose and Fermi statistics are incorporated correctly. Note that there is no
nonabelian gauge symmetry. Therefore, unlike Yang--Mills theories, there are no additional
group-theory factors. In this respect Born--Infeld theories and
world-volume theories of single branes are more like gravitational theories.

\subsection{Poincar\'e and little-group invariance}

The M5-brane theory is defined on a 6d Minkowski
hypersurface embedded in an 11d Minkowski spacetime. Therefore Spin$(5,1) \times {\rm Spin}(5)$
is the unbroken subgroup of the 11d Lorentz group Spin$(10,1)$. The six translation symmetries
along the brane are preserved, whereas the five transverse to the brane are spontaneously
broken. Indeed, the spectrum includes five massless scalars, which are the corresponding
Goldstone bosons.

The 6d Lorentz group, Spin$(5,1)$, is a noncompact version of
SU$(4)$. Like SU$(4)$, it has two inequivalent four-dimensional spinor representations. In the case
of Spin$(5,1)$ they have opposite chirality. Their components are represented
by indices $A,B = 1,2,3,4$, written as superscripts for the ``left-handed" representation
and as subscripts for the ``right-handed" one. Invariant tensors are $\e_{ABCD}$,
$\e^{ABCD}$, and $\d_A^B$. The last one is used to contract superscripts with subscripts.
The momentum six-vector can be expressed as an antisymmetric matrix, $P_{AB} = \s^{\mu}_{AB} P_{\mu}$
or $P^{AB} = \half \e^{ABCD} P_{CD}$.
Translation invariance leads to momentum conservation $P^{AB} =\sum_{i=1}^n p_i^{AB} =0$.
When we discuss amplitudes it will sometimes be useful to extract the momentum-conservation
delta function and to define
\be
{\cal A}_n = \d^6(P^{AB}) A_n.
\ee

In addition, the M5-brane theory has a USp$(4) ={\rm Spin}(5)$ R-symmetry group.
As implied above,
this symmetry arises from invariance of the brane under rotations of the five M-theory directions
transverse to the M5-brane. By definition, this is an R symmetry, because the supersymmetry
charges belong to nontrivial representations of this group. R symmetry and supersymmetry will be described
further in the subsequent subsections.

Massless particles in 6d have on-shell degrees of freedom
classified by representations of the little group ${\rm Spin}(4) = {\rm SU}(2)_L \times
{\rm SU}(2)_R$. There is a distinct such group for each particle. Each M5-brane excitation is
described by a $(2,0)$ tensor supermultiplet, which is a singlet of its SU$(2)_R$.
Therefore, SU$(2)_R$ symmetry is trivial in the M5-brane theory,
and we will not refer to SU$(2)_R$ any more in this work.\footnote{This property is
not shared by the other 6d theories that we have studied. For this reason their amplitudes are
somewhat more complicated. Also, super Yang--Mills and supergravity theories
have nonvanishing amplitudes for odd $n$, which
is another complicating feature.} Thus, on-shell particles are characterized by representations of the R-symmetry group USp$(4)$ and the little group SU$(2)_L$.
Note that there is just one USp$(4)$ R-symmetry group, which classifies all $n$ of the particles,
but there are $n$ separate SU$(2)_L$ little groups -- one for each particle.

Indices $a,b,\ldots = +,-$
are used to label doublets of SU$(2)_L$, but such labels are meaningless unless they are accompanied by a label $i,j,\dots = 1,2, \ldots n$, so that we know which particle's little group they refer to.
(We prefer not to use the more cumbersome notation $a_i, b_i, \ldots$.)
Invariant tensors (for a given little group) are $\e^{ab}$ and $\e_{ab}$. The spectrum of the
M5-brane theory consists of a single
$(2,0)$ tensor supermultiplet, which contains the representations $({\bf 5,1}) + ({\bf 4,2}) + ({\bf 1,3})$
of USp$(4) \times {\rm SU}(2)_L$. For example, the third term is the chiral tensor, denoted $\b^{ab} = \b^{ba}$. In a covariant Lagrangian formalism a chiral tensor is
described by a two-form with a self-dual field strength and the appropriate gauge symmetry.
This formalism introduces complications that are circumvented when on-shell amplitudes
are described using the spinor-helicity formalism that we will now describe.
A $(2,0)$ tensor multiplet is conveniently expressed as a scalar function of four Grassmann parameters
$\eta_a^I$, where $I=1,2$. It is easy to verify that it contains the SU$(2)_L$ representations listed
above with the correct multiplicities. The way that USp$(4)$ is realized will be explained later.

The only way that non-trivial Lorentz-group representations appear in our description
of M5-brane amplitudes is through spinor-helicity coordinates $\l^A_{ia}$ \cite{Cheung:2009dc}.
The basic idea is that the momentum of the $i$-th particle is given by
\be
p_i^{AB} = \e^{ab}\l_{ia}^A \l_{ib}^B,
\ee
which is invariant under the $i$-th little group (and all the others). Because of this,
three of the eight components of $\l_i$
are redundant. The remaining five encode the five independent components of the momentum vector
of a massless particle in 6d. (The mass squared is proportional to the Pfaffian
of $p_i^{AB}$, which vanishes because the matrix only has rank two.)
The $15$ generators of the 6d Lorentz group are given by the traceless matrix
\be
J^A_B = \l^A \cdot \frac{\pa}{\pa\l^B} -\frac{1}{4}\d^A_B \l^C \cdot \frac{\pa}{\pa\l^C}.
\ee
The index contraction implied by the dots in this formula corresponds to the tensor $\d^{ia}_{jb}$.
The rest of the Poincar\'e group is generated by the total momentum $P^{AB} = \l^A \cdot \l^B$.
In this case the dot corresponds to $\d^{ij}\e^{ab}$.
We could also define generators of conformal transformations
\be
K_{AB} = \frac{\pa}{\pa\l^A} \cdot \frac{\pa}{\pa\l^B} ,
\ee
though the M5-brane theory does not have conformal symmetry.  Similarly, the dilatation operator
\be
D = \l^A \cdot \frac{\pa}{\pa\l^A}
\ee
also does not generate a symmetry. However, it is useful to know that $(D-2n) {A}_n =0$,
which implies that ${A}_n \sim \l^{2n}$.

The SU$(2)_L$ little group of the $i$-th particle is generated by the traceless operator
\be
j_{ia}^b = \eta_{ia}^I \frac{\pa}{\pa\eta^I_{ib}} - \half \d^b_a \, \eta_{ic}^I\frac{\pa}{\pa\eta^I_{ic}}.
\ee
The sum of the trace terms,
\be
d = \eta^I \cdot \frac{\pa}{\pa\eta^I},
\ee
is analogous to $D$. It even has the same eigenvalue, $(d-2n) {A}_n =0$. Thus,
$A_n \sim \eta^{2n}$.

\subsection{R symmetry}

As explained in the previous subsection, the M5-brane theory has USp$(4)$ R symmetry.
However, since it is convenient to use the Grassmann parameters introduced in the
previous subsection, only an ${\rm SU}(2)\times {\rm U}(1)$ subgroup is manifest.
The SU$(2)$ factor is generated by
\be
R^I_J = \eta^I \cdot \frac{\pa}{\pa\eta^J} -\half \d_J^I\, \eta^K \cdot \frac{\pa}{\pa\eta^K} .
\ee
The U$(1)$ subgroup of the R-symmetry group, identified at the end of the previous subsection,
is generated by
\be
R = d-2n  = \half \left(\eta^I \cdot \frac{\pa}{\pa\eta^I}
- \frac{\pa}{\pa\eta^I} \cdot \eta^I \right).
\ee
Thus, the requirement that $d=2n$ is a consequence of this U$(1)$ subgroup
of the R symmetry.

The remaining six R symmetries will not be manifest, though we will explain how to prove that they
are symmetries of the amplitudes. These generators are given by a pair of symmetric matrices:
\be
R^{IJ} = \eta^I \cdot \eta^J \quad {\rm and } \quad
R_{IJ} = \frac{\pa}{\pa\eta^I} \cdot \frac{\pa}{\pa\eta^J}.
\ee
It is easy to verify that these 10 generators give the USp$(4)$ Lie algebra.
One way to think about this is that generators $R^{\tilde I \tilde J} = R^{\tilde J \tilde I}$,
where $\tilde I$ and $\tilde J$ take values from 1 to 4, can be obtained by
replacing {\em subscripts} $I,J$ by {\em superscripts} $I+2, J+2$. Thus $R^1_1  \to R^{13}$,
$R_{12} \to R^{34}$, etc. This corresponds to raising an index using the
antisymmetric symplectic metric $\o^{\tilde I \tilde J}$ with
$\o^{13} = \o^{24} =1$. In this notation, the full R symmetry algebra is
\be
[R^{\tilde I \tilde J}, R^{\tilde K \tilde L}] = \o^{\tilde I \tilde K} R^{\tilde J \tilde L}
+ \o^{\tilde J \tilde K} R^{\tilde I \tilde L}
 + \o^{\tilde I \tilde L} R^{\tilde J \tilde K} + \o^{\tilde J \tilde L} R^{\tilde I \tilde K} .
\ee
So the algebra is easy enough to understand. What is more challenging is
to ensure that the amplitudes actually possess the symmetries generated by $R^{IJ}$ and $R_{IJ}$.

As was already mentioned, the supermultiplet $\Phi(\eta)$ contains the
representation $({\bf 5,1}) + ({\bf 4,2}) + ({\bf 1,3})$ of USp$(4) \times {\rm SU}(2)_L$.
In the expansion of $\Phi(\eta)$ in powers of $\eta$ the five scalars appear in three
different terms, even though they form an irreducible R-symmetry multiplet.
Similarly, the spinors appear in two terms.  The five scalars
can be described by an antisymmetric matrix $\phi^{\tilde I \tilde J}$ with a vanishing
symplectic trace ($\o_{\tilde I \tilde J} \phi^{\tilde I \tilde J} = 0$).\footnote{Equivalently,
one could utilize a
five-vector $\phi^\a = \half \s^{\a}_{\tilde I \tilde J} \phi^{\tilde I \tilde J} $.}
Similarly, the four little-group spinors
are denoted $\psi^{a\tilde I}$ and the little-group triplet remains $\b^{ab}$ as before.
Writing these indices as superscripts is a matter of convention, since they can be lowered using
the symplectic metrics $\e_{ab}$ and $\omega_{\tilde I \tilde J}$.

\subsection{Supersymmetry}

M theory in an 11d Minkowski background
has 32 conserved supercharges, and only half of them preserve a flat 6D hypersurface.
Therefore the M5-brane theory has 16 conserved supercharges, which transform as
$({\bf 4,4})$ with respect to ${\rm Spin}(5,1) \times {\rm USp}(4)$.  Anticommutators of
the broken supercharges give momenta that generate five transverse translations of the M5-brane,
which are also broken symmetries. Like the scalars,
the massless fermions of the M5-brane theory can be interpreted as Goldstone particles
associated to the spontaneously broken symmetries.

Eight anticommuting supercharges are given by
\be
Q^{AI} = \sum_{i=1}^n q_i ^{AI} = \l^A \cdot \eta^I
\quad {\rm where} \quad
{q}_i^{AI} = \e^{ab} \l^A_{ia} \eta^I_{ib}.
\ee
Their conservation can be implemented by Grassmann delta functions:
\be
A_n(\l, \eta) = \d^8(Q^{AI}) F_n(\l,\eta).
\ee
The other eight supercharges, which are also mutually anticommuting, are represented by
\be
{\overline Q}^A_I = \sum_{i=1}^n {\bar q}^A_{iI}
= \l^A \cdot \frac{\partial}{\partial \eta^I},
\quad {\rm where} \quad
{\bar q}^A_{iI} = \l^A_{ia} \frac{\partial}{\partial \eta^I_{ia}}.
\ee
The nonzero supersymmetry anticommutators are
\be
\{ Q^{AI}, {\overline Q}^B_J \} = \d^I_J P^{AB}.
\ee
The entire supersymmetry algebra can be combined into the
single equation
\be
\{ Q^{A \tilde I}, Q^{B \tilde J} \} = \o^{\tilde I \tilde J} P^{AB}.
\ee
Because of the appearance of the factor $\d^8(Q^{AI})$ in the amplitudes, half of the
supersymmetry is manifest, and the other half needs to be proved. Also, only a subgroup of the
USp$(4)$ R symmetry is manifest. The approach taken in \cite{Heydeman:2017yww} is to explicitly verify
the rest of the R symmetry of the $n$-particle amplitudes,
since this together with $\d^8(Q^{AI})$ implies the rest of the supersymmetry.

\subsection{R symmetry of the four-particle amplitude}

The four-particle amplitude of the M5-brane theory is given (up to a constant)
by the deceptively simple formula
\be \label{A4}
A_4 = \d^8( Q^{AI} ).
\ee
As before, $Q^{AI} = \sum_{i=1}^4 \e^{ab} \l^A_{ia} \eta^I_{ib} $. The formula (\ref{A4}) makes half
of the supersymmetry manifest. Its main deficiency is that
the USp$(4)$ R symmetry is very obscure. The point is that the index $I=1,2$ labels a doublet
of an SU$(2)$ subgroup of the R symmetry. As mentioned earlier, the symmetry associated to
the six generators $R^{IJ}$ and $R_{IJ}$ is far from obvious. The solution to this problem for $n=4$,
presented in \cite{Heydeman:2017yww}, is repeated here. The generalization to all $n$ is also
given in \cite{Heydeman:2017yww}.

For this purpose it is convenient to rename
$\eta_{i-}^I$ as $\eta_{i}^I$ and $\eta_{i+}^I$ as
$\tilde\eta_{i}^I$. Then we Fourier transform the latter coordinates
to conjugate Grassmann coordinates denoted $\z_{iI}$. Thus, we consider
\begin{equation}
\tilde A_4 = \int d^8\tilde\eta^I_i e^{\sum_{iI}\tilde\eta^I_i \z_{iI}}
\d^8 \left(\sum_{i=1}^4 \e^{ab} \l^A_{ia} \eta^I_{ib}  \right) .
\end{equation}
Substituting an integral representation of the delta functions,
\be
\d^8( Q^{AI} ) = \int d^8\th_{AI} e^{\th_{AI} Q^{AI}},
\ee
and carrying out the $\tilde\eta$ integrations gives
\begin{equation}
\tilde A_4 = \int d^8 \th_{AI} \d^8(\z_{iI}
+ \sum_A \th_{AI} \l^A_{i-} )\, e^{\sum_{AIi} \th_{AI} \l^A_{i+} \eta^I_i}.
\end{equation}
If the $4\times 4$ matrix $\l^A_{i-}$ is nonsingular, which is
generically the case, then
\begin{equation}
\d^8(\z_{iI} + \sum_A \th_{AI} \l^A_{i-} )
 = (\det \l_-)^2 \d^8((\z \l_-^{-1})_{IA} +  \th_{AI} ),
\end{equation}
where $(\z \l_-^{-1})_{IA} = \sum_i \z_{iI} (\l_-^{-1})_{iA}$. Thus.
\begin{equation}
\tilde A_4 = (\det \l_-)^2  \exp(-\tr (\z \l_-^{-1}\l_+\eta)).
\end{equation}
More explicitly, the exponent is
\begin{equation}
-\tr (\z \l_-^{-1}\l_+\eta) = \tr(\l_-^{-1}\l_+\eta \z))
= \sum_{ij} (\l_-^{-1}\l_+ )_{ij} (\eta\z)_{ji} ,
\end{equation}
where $(\eta\z)_{ji} = \eta_j^I \z_{iI}$.

Momentum conservation implies that $(\l_+ \l_-^T)^{AB} = (\l_- \l_+^T )^{AB}$,
and therefore $(\l_-^{-1}\l_+ )_{ij}$ is a symmetric matrix. Since
only the symmetric part of $(\eta\z)_{ji}$ contributes, it can be replaced by half of
\begin{equation}
E_{ij} = \sum_{I=1}^2 \left(\eta^I_i \z_{Ij} +  \eta^I_j \z_{Ii} \right ).
\end{equation}
$E$ can now be rewritten in a form with manifest USp$(4)$ R symmetry
\begin{equation}
E_{ij} = \sum_{\tilde I, \tilde J=1}^4 \o_{\tilde I \tilde J}\eta^{\tilde I}_i \eta^{\tilde J}_j ,
\end{equation}
where we have renamed $\z_{Ii} = \eta^{I+2}_i$. As before, the only nonzero elements of the symplectic
metric $\o_{\tilde I \tilde J}$ are $\o_{13} = \o_{24} = -\o_{31} = - \o_{42} = 1$.
Then $\eta^{\tilde I}_i$, with ${\tilde I}=1,2,3,4$, belongs to the fundamental representation of the
USp$(4)$ R-symmetry group. To summarize, we have shown that $\tilde A_4$ can be written
in the manifestly R-symmetric form
\begin{equation}
\tilde A_4 = \D_4  e^{-\half\sum_{ij} (K_4)_{ij} E_{ij}},
\end{equation}
where
\be
\D_4 = (\det \l_-)^2 \quad \and \quad (K_4)_{ij} = (\l_-^{-1}\l_+)_{ij}.
\ee
$A_4$ can be recovered as the inverse Grassmann Fourier transform. In conclusion,
USp$(4)$ R symmetry is a property of ${\tilde A}_4$, and not $A_4$, which is good enough.
However, the formula for $A_4$ is simpler and makes the supersymmetry
more transparent.

\subsection{Symplectic Grassmannian}

The $n$-particle M5-brane superamplitude can be written in a way that depends on
a symplectic Grassmannian, $\mathbb{LG}(n, 2n)$, which has ${\rm USp}(2n)$
symmetry \cite{Schwarz:2019aat}. The way
this works is that the subscripts $ia$ on the $\l$'s and $\eta$'s are combined
to label the fundamental $2n$-dimensional representation of USp$(2n)$. USp$(2)^n$,
the product of the $n$ little groups, each of which is SU$(2)_L = {\rm USp}(2)$,
is a subgroup of this USp$(2n)$.\footnote{Other analogous
examples include the D3-brane theory in which U$(1)^n$ is enhanced to U$(n)$
and the M2-brane theory in which ${\bf Z}_2^n$ is enhanced to O$(n)$.}
This combines and extends the little-group symmetries and the permutation symmetry
of the amplitude.

The symmetry appears in the form of a symplectic Grassmannian, denoted
$\mathbb{LG}(n, 2n)$, which is a homogeneous space of $n(n+1)/2$
complex dimensions. One description of this space is USp$(2n)$/ U$(n)$.
In particular, the four-particle amplitude is given (up to a constant) by
${\cal A}_4 = \d^6(P^{AB}) \d^8(Q^{AI})$. Since $P^{AB} = \l^A \cdot \l^B$ and
$Q^{AI} = \l^A \cdot \eta^I$, the USp$(8)$ symmetry of this amplitude is manifest.
Also, these factors generalize to all $n$, though only even $n$ is relevant.

\section{${\mathbf n}$-particle M5-brane amplitudes}

In the spirit of the CHY construction of $n$-particle scattering amplitudes
\cite{Cachazo:2014xea}\cite{Cachazo:2013hca},
the formulas can be schematically summarized in the following form,
\be
\mathcal{A}_n = \int d\mu^{\rm 6D}_n \, \mathcal{I}_L \, \mathcal{I}_R \, ,
\ee
where the measure $d\mu^{\rm 6D}_n$, which is theory independent, encodes the 6d
massless kinematics including momentum conservation.
It provides a map from 6d kinematics to punctures of the Riemann sphere.
The factors $\mathcal{I}_L$ and $\mathcal{I}_R$ determine the specific theory under consideration.

A coordinate $\sigma_i$ is assigned to each of the $n$ massless external particles.
These coordinates are defined up to an overall common
$\slc_\sigma$ M{\"o}bius-group transformation,
\bea
\sigma_i \rightarrow \frac{a\sigma_i + b}{c\sigma_i + d}, \qquad ad-bc = 1.
\eea
This allows the coordinates of three of the punctures to be given arbitrary distinct values.
The contribution to the integration measure,
$\prod_{i=1}^n d \sigma_i$/vol($\slc_\sigma$), is defined in a standard way.

A function of the $\s$ coordinates $F(\{ \sigma_i \} )$ is said to have weight $w$ if it
transforms under a M{\"o}bius transformation by the rule
\bea
F \left( \left\{ \frac{ a\sigma_i +b} {c\sigma_i +d }\right\} \right)
= \left[ \prod_{i=1}^n(c\sigma_i + d)\right]^w F(\{ \sigma_i \} ).
\eea
The measure $d\mu^{\rm 6D}_n$, defined below, transforms with weight $w=-4$.
Therefore, $\mathcal{I}_L \, \mathcal{I}_R$ must have weight 4. In practice,
each of the two factors, $\mathcal{I}_L$ and $\mathcal{I}_R$, has weight 2.

The presentation that follows is based on \cite{Schwarz:2019aat}.
For even $n$, which is all that is required for the M5-brane theory, the $n$-particle measure is given by
\bea \label{eq:rationalMap}
\prod_{i=1}^n \delta(p_i^2)\int d\mu^{\rm 6D}_n =\int \frac{\prod_{i=1}^n d\sigma_i\, \prod_{k=0}^{m} d^8
\rho_k}{\vol( \slc_\sigma \times \slc_\rho)} \frac{1}{V_n^2}\prod_{i=1}^n  \delta^6 \left( p^{AB}_i -
\frac{\langle \rho^{A}(\sigma_i)\, \rho^{B}(\sigma_i) \rangle }{\prod_{j\neq i} \sigma_{ij}}\right)\, ,
\eea
where
\bea
n=2m+2, \quad \sigma_{ij} =\s_i - \s_j, \quad {\rm and} \quad V_n = \prod_{i<j} \s_{ij} \, .
\eea
The mass-shell delta functions allow the momenta $p^{AB}_i$ to be expressed in terms of
spinor-helicity coordinates $\l^A_{ia}$.
The maps in the delta functions are given by degree-$m$ polynomials,
\bea
\rho^{A}_a(\sigma) = \sum_{k=0}^{m} \rho^A_{a,k}\, \sigma^k \, .
\eea
They are determined up to an overall $\slc_\rho$ transformation, which is a complexification of ${\rm SU}(2)_L$, and its volume also is divided out.
The $\slc_\sigma$ transformations of the coordinates $\rho^A_{a,k}$ are determined by requiring that the expressions inside the delta functions in (\ref{eq:rationalMap}) are invariant. Then one can show that $d\mu^{\rm 6D}_n$ has weight $-4$.

As shown in~\cite{Cachazo:2018hqa}, by introducing $n$ additional $2 \times 2$ matrices $(W_i)^b_a$, the dependence on the $\rho$ coordinates can be recast in the linear form
\begin{align} \label{eq:linear-1}
\int \! d\mu_{n}^{\text{6D}}  = \int\! \frac{\prod_{i=1}^n\, d\sigma_i\,  d^{4}W_i \prod_{k=0}^{m} d^8
\rho_k}{\vol( \slc_\sigma \times \slc_\rho)}  \prod_{i=1}^n \d^8 \!\left ( \l^A_{ia} - (W_i)^b_a \r^A_{b}
(\s_i) \right ) \d\!\left(|W_i| - \frac{1}{\prod_{j\neq i}\sigma_{ij} } \right) \, ,
\end{align}
where $|W_i|={\rm det}\, W_i$.
The indices $a$ and $b$ of $(W_i)^b_a$ refer to different groups. Specifically, $b$
is contracted with the $\slc$ index of the moduli $\rho^A_{k, b}$, and therefore
it is a global index, whereas $a$ is associated with the little group of
the $i$-th particle.

Integrating out the moduli $\rho^A_{a,k}$ \cite{Cachazo:2013zc} leaves
\begin{align} \label{eq:emerge-Gr}
\int \! d\mu_{n}^{\text{6D}} =
 \int\! \frac{\prod_{i=1}^n d\sigma_i\, d^{4}W_i  }{\vol( \slc_\sigma \times \slc_W)} \prod^m_{k=0}
 \d^{2\times 4} \left(\sum_{i=1}^n (W_i)^b_a \s^k_i \l^{A a}_{i}\right)\prod_{i=1}^n \d\left(|W_i|
 - \frac{1}{\prod_{j\neq i}\sigma_{ij} }\right) \, ,
\end{align}
where $\slc_\rho$ has become $\slc_W$, which is the symmetry acting on the global
little-group index $b$. We can now show the emergence of the symplectic Grassmannian by
defining the $n\times 2n$ matrix
\begin{equation} \label{eq:C-matrix}
{C}_{k, b;i, a}=(W_i)^b_a\, \s_i^k\, .
\end{equation}
We have grouped the exponent $k$ with the global $\slc$ index $b$ to label $n$ rows
and the index $i$ with the $i$-th little-group index $a$ to label $2n$ columns.
The matrix $C$ formed in this way satisfies the identity
\begin{align} \label{eq:identity}
C\cdot \Omega \cdot C^T =0 \,,
\end{align}
where $\Omega$ is the USp$(2n)$ metric
\begin{equation}
\Omega= \begin{pmatrix}
0 & \mathbb{I}_n \\
- \mathbb{I}_n  & 0
\end{pmatrix} \, ,
\end{equation}
and $\mathbb{I}_n$ is the $n \times n$ identity matrix. (\ref{eq:identity})
is proved by using the delta-function constraints and the theorem\footnote{This theorem
is easy to establish by showing that the
residues of the poles vanish when $K$ is a non-negative integer and that the expression vanishes
at infinity for $K < n-1$.}
\begin{equation}\label{eq:sigforms}
\sum_{i=1}^n \frac{\sigma_i^K}{\prod_{j\neq i} \s_{ij}} =0 \quad {\rm for }\quad  K=0,1,\ldots, n-2\, .
\end{equation}
If $M$ is a symplectic matrix belonging to ${\rm USp}(2n)$, satisfying
$M^T \cdot \Omega\cdot M = \Omega$, the identity (\ref{eq:identity}) is invariant under
the symplectic transformation $C\to C\cdot M^T$.
The scattering-equation constraints can then be encoded as
\bea \label{eq:RMap-linear}
 \prod_{k=0}^{m} \delta^{2\times 4} \left(\sum_{i=1}^n {C}_{k, b;i, a} \, {\lambda}^{A a}_i \right)
= \delta^{n\times 4} (C \cdot \Omega \cdot \Lambda^A ) \, ,
\eea
with $C \cdot \Omega \cdot C^T=0$. We have introduced a $2n$-dimensional vector $\Lambda^A$, which is also
a Lorentz spinor, built out of the spinor-helicity coordinates $\lambda^{A}_{i, a}$,
\begin{equation}
\Lambda^A :=\{ \lambda^{A}_{1, 1}, \lambda^{A}_{2, 1}, \ldots, \lambda^{A}_{n, 1},  \lambda^{A}_{1, 2},
\lambda^{A}_{2, 2}, \ldots, \lambda^{A}_{n, 2}\}  \,  .
\end{equation}
Invariance under symplectic transformations requires that $\Lambda^A \rightarrow M \cdot \Lambda^A$.

Let us verify that $C$ parametrizes $\mathbb{LG}(n,2n)$. First of all, it has
USp$(2n)$ symmetry, as required. We can also check the complex dimension of the space that
it parametrizes, which is supposed to be $n(n+1)/2$.
$C$ is an $n \times 2n$ complex matrix, which has $2n^2$ complex dimensions. However,
it can be multiplied on the left by an arbitrary ${\rm GL}(n, \IC)$ matrix,
without changing the constraint equations $C \cdot \Omega \cdot C^T =0$.
Altogether, we are left with $2n^2 - n^2 - n(n-1)/2 = n(n+1)/2$ complex dimensions, as required.
These two facts uniquely characterize $\mathbb{LG}(n,2n)$. ${\rm GL}(n, \IC)$ transformations
also preserve the scattering equations $C \cdot \Omega \cdot \Lambda^A =0$

The formula that describes
the tree amplitudes of the M5-brane theory is given by~\cite{Schwarz:2019aat}
\bea \label{eq:M5-formula}
\mathcal{A}^{\rm M5}_{n} =
 \int d \mu_n^{\rm 6D}  \, {\cal I}_L^{(2,0)} \,  {\cal I}_R^{\rm DBI} \, ,
\eea
where the factors ${\cal I}_L^{(2,0)}$ and ${\cal I}_R^{\rm DBI}$ in the integrand are
\bea \label{eq:M5-ILIR}
{\cal I}_L^{(2,0)}
= \delta^{n\times 2} (C \cdot \Omega \cdot \eta^I ) \, V_n\,  { {\rm Pf}^{\prime} S_n  } \, , \qquad
{\cal I}_R^{\rm DBI}
=  \left( {\rm Pf}^{\prime} S_n \right)^2 =  {\rm det}^{\prime} S_n\, .
\eea
$S_n$ is an $n\times n$ antisymmetric matrix that has rank $n{-}2$ and is given by
\bea \label{eq:sij}
[S_n]_{ij} = \frac{ p_i \cdot p_j}{\sigma_{ij}}\, .
\eea
The reduced Pfaffian of  $S_n$ is defined as
\bea \label{eq:Sn-matrix}
{\rm Pf}^{\prime} S_n = \frac{(-1)^{k+l}}{\sigma_{kl}} {\rm Pf} (S_n)^{kl}_{kl} \, ,
\eea
where $(S_n)^{kl}_{kl}$ is an $(n{-}2) \times (n{-}2)$ matrix with the $k$-th and $l$-th rows
and columns of $S_n$ removed, and the result is independent of the choice of $k, l$.
Since ${\rm Pf}' S_n$ has conformal weight $w=1$, we see that ${\cal I}_R^{\rm DBI}$
has conformal weight $w=2$, as required. This factor, which is only nonzero
for even $n$, appears in all DBI-type theories. These theories only have nonvanishing
amplitudes when $n$ is even. The factor ${\cal I}_L^{(2,0)}$ also has weight 2,
since $\delta^{n\times 2} (C \cdot \Omega \cdot \eta^I )$ has weight $n$ and $V_n$
has weight $1-n$. As the notation suggests, it implements $(2,0)$ supersymmetry. The
fact that the residues of the poles in these amplitudes have the factorization
properties required by perturbative unitarity has been established 
in \cite{Albonico:2020mge}.

In conclusion, the M5-brane theory has an interesting mix of symmetries, and its tree amplitudes
can be written in a form that enables one to understand all of them. The form of the scattering
equations described in this paper is sometimes referred to as the ``rational maps" approach.
An alternative, based on ``polarized scattering equations,'' has been proposed by
Geyer and Mason \cite{Geyer:2018xgb}. One of the main results of \cite{Schwarz:2019aat}
is to show that the two descriptions are related by a
${\rm GL}(n, \mathbb{C})$ transformation of $C$. A possible goal for the future is to
extend this type of analysis to string theory and M-theory amplitudes.

I wish to acknowledge my collaborators Matthew Heydeman and Congkao Wen
for their important contributions to this work.
This research has been supported in part by the Walter Burke Institute
for Theoretical Physics at Caltech and by U.S. DOE Grant DE-SC0011632,
and some of this work was performed at the Aspen Center for Physics,
which is supported by National Science Foundation grant PHY-1607611.


\end{document}